\documentclass[prb,twocolumn,showpacs]{revtex4}

\newif\ifpdf
	\ifx\pdfoutput\undefined
	\pdffalse 
	\else
	\pdfoutput=1 
	\pdftrue
	\fi

	\ifpdf
	\usepackage[pdftex]{graphicx}
	\else
	\usepackage{graphicx}
	\fi

\begin{document}
\title{Specific heat and $\mu^+$SR measurements 
in Gd(hfac)$_3$NITiPr molecular magnetic chains:
indications for a chiral phase without long range helical order}

\author{A. Lascialfari$^1$}
\email{lascialfari@fisicavolta.unipv.it}
\author{R. Ullu$^1$}
\author{M. Affronte$^2$}
\email{affronte@unimo.it}
\author{F. Cinti$^2$}
\author{A. Caneschi$^3$}
\author{D. Gatteschi$^3$}
\author{D. Rovai$^3$}
\author{M.G. Pini$^4$}
\author{A. Rettori$^5$}

\affiliation{
$^1$ I.N.F.M. and Dipartimento di Fisica ``A.Volta", Universit\`a di Pavia, 
Via Bassi 6, I-27100 Pavia, Italy 
\\
$^2$ I.N.F.M. - S$^{3}$ National Research Center and Dipartimento di Fisica, 
Universit\`a di Modena e Reggio Emilia, Via G.Campi 213/A, I-41100 Modena, Italy 
\\
$^3$ I.N.S.T.M. and Dipartimento di Chimica, Universit\`a di Firenze,
Via della Lastruccia 3, I-50019 Sesto Fiorentino (FI), Italy
\\
$^4$Istituto di Fisica Applicata ``Nello Carrara",
Consiglio Nazionale delle Ricerche, Via Panciatichi 56/30,
I-50127 Firenze, Italy 
\\
$^5$I.N.F.M. and Dipartimento di Fisica, Universit\`a di Firenze,
Via G. Sansone 1, I-50019 Sesto Fiorentino (FI), Italy
}

\date{\today}

\begin{abstract}

Low temperature specific heat $C(T)$ and zero-field muon spin 
resonance ($\mu^+$SR) measurements were performed in 
Gd(hfac)$_3$NITiPr, a quasi one-dimensional molecular magnet 
with competing nearest neighbor and next-nearest neighbor 
intrachain exchange interactions. The specific  heat data exhibit 
a $\lambda$-peak at $T_0$=2.08$\pm$0.01K that disappears 
upon the application of a 5 Tesla magnetic field. 
Conversely, the $\mu^+$SR data do not present any anomaly at 
$T\approx 2$ K, proving the lack of divergence of the two-spin 
correlation function as required for usual three-dimensional 
(3D) long range helical order. Moreover, 
no muon spin precession can be evinced 
from the $\mu^+$SR asimmetry curves, 
thus excluding the
presence of a long range ordered magnetic lattice.
These results provide indications 
for a low $T$ phase where chiral order is established in absence 
of long range helical order. 
\end{abstract}


\maketitle

\section{Introduction}
For three-dimensional (3D) $XY$ helimagnets, Kawamura\cite{Kawamura} 
proposed a new universality class as a consequence of frustration. 
In addition to the SO(2) symmetry of the spin variable 
${\bf S}_{\bf r}$, there is a further degree of freedom, 
the spin chirality,\cite{Villain} defined as 
$\kappa_{12}=[{\bf S}_{{\bf r}_1} \times 
{\bf S}_{{\bf r}_2}]^z/ \vert\sin(Qa)\vert$
(${\bf Q}$ is the pitch of the helical ground state,
${\bf S}_{{\bf r}_1}$ and ${\bf S}_{{\bf r}_2}$ are spins on 
nearest neighbor planes perpendicular to ${\bf Q}$). For $T=0$, the 
order parameter $\kappa_{12}=\pm 1$ describes the clockwise 
or anti-clockwise degeneracy Z$_2$ of the helical structure. 
If chiral order and spin order occur simultaneously,
the order parameter space becomes Z$_2 \times$ SO(2), leading to
a new universality class.

For quasi one-dimensional (1D) helimagnets, 
a different behavior was predicted by Villain.\cite{Statphys}
In addition to the low temperature ($T<T_N$) 3D long range helical 
phase and to the high temperature ($T>T_{0}$) paramagnetic one, 
Villain proposed that a chiral phase is stable for $T_N<T<T_{0}$. 
The chirality order parameter associated to a nearest 
neighbor spin pair is 
$\kappa_n=[{\bf S}_n \times {\bf S}_{n+1}]^z/|\sin(Qa)|$.
The chiral phase can be described as a collection of corkscrews 
that turn all clockwise or all anti-clockwise, whereas their phases 
are random.\cite{Statphys} Due to the Ising nature of chirality, 
$T_{0}$ results greater than $T_N$ because the chirality-chirality
correlation function is stronger than the usual spin-spin correlation.
In fact the chiral correlation length $\xi_{\kappa}$, defined via 
the chirality-chirality correlation function  
$\langle \kappa_{1}\kappa_{n+1} 
\rangle \approx A \exp(-na/\xi_{\kappa})$,
diverges exponentially as temperature is decreased 
$\xi_{\kappa} \approx \exp(|J|/T)$ (where $J$ is the exchange
constant), while the spin correlation length $\xi_s$, 
defined via the spin-spin correlation function 
$\langle {\bf S}_{1} \cdot {\bf S}_{n+1} \rangle \approx
A^{\prime} \exp(-na/\xi_s)$, diverges as a power law
$\xi_s \approx |J|/T$.\cite{Kawa} As a consequence,  
an anomaly is expected at  $T_{0}$ 
for the physical properties related to 
$\langle \kappa_{1} \kappa_{n+1} \rangle$, such as the 
magnetic specific heat, while the quantities related to 
$\langle {\bf S}_{1} \cdot {\bf S}_{n+1} \rangle$, 
such as the magnetic susceptibility, are expected to 
behave critically only at $T_N$. At $T_0$, a
continuous phase transition or a first-order one can be 
present.\cite{Statphys}

Villain's prediction\cite{Statphys} was tested in the 
molecular-based quasi-1D magnet 
Gd(hfac)$_3$NITiPr\cite{Bartolome,Affronte}
[hfac is hexafluoro-acetylacetonate and NITiPr is 
2-isoPropyl-4,4,5,5-tetramethyl-4,5-dihydro-1H-imidazolyl-1-oxyl 
3-oxide]; 
iPr are isoPropyl organic radicals (with spin $s=1/2$)
that alternate with rare-earth Gd$^{3+}$ magnetic ions (with spin 
$S=7/2$) along the chain direction, $z$. The spins interact through 
competing nearest-neighbor ($J_1>0$) and next nearest-neighbor 
($J_2<0$ and $J_2^{\prime}<0$) exchange constants along the chain 
\begin{eqnarray}
&{\cal H}&_{i,{\rm intra}}= 
\sum_{n=1}^{N/2}
\bigg\{
-J_1 ({\bf S}_{i,2n-1}\cdot{\bf s}_{i,2n}+
{\bf s}_{i,2n}\cdot{\bf S}_{i,2n+1})
\cr
&-&J_2({\bf S}_{i,2n-1}\cdot{\bf S}_{i,2n+1})
+D(S^z_{i,2n-1})^2-g\mu_B {\bf H}\cdot{\bf S}_{i,2n-1}
\cr
&-&J_2^{\prime}({\bf s}_{i,2n}\cdot {\bf s}_{i,2n+2})
-g^{\prime}\mu_B{\bf H}\cdot{\bf s}_{i,2n}
\bigg\}
\end{eqnarray}
where the index $n$ refers to the spin position along 
the $i$-th chain. $D>0$ is an effective 
anisotropy favouring the spins to lie in the $xy$ plane
(perpendicular to the chain axis, $z$). 
$H$ is an external magnetic field and $g,g^{\prime}$ are the
gyromagnetic factors of the spins $S$ and $s$, respectively.
The next-nearest  neighbor antiferromagnetic exchange is 
dominant, leading to a helical ground state with pitch
$\pm Qa=\pm \cos^{-1}\left\lbrack 
{1\over { 2 (\delta+\delta^{\prime}) }} \right\rbrack$, 
where $\delta=(\vert J_2\vert S^2)/(J_1 s S)$,
$\delta^{\prime}=(\vert J_2^{\prime}\vert s^2)/(J_1 s S)$
and $2(\delta+\delta^{\prime})>1$.
Different chains interact through a weak ferromagnetic 
interchain exchange ($0<J_{\perp}<<J_1$)
\begin{equation}
{\cal H}_{\rm inter}=\sum_{\langle i \ne j \rangle}
\sum_{n=1}^{N}\left\lbrack
-J_{\perp} ({\bf S}_{i,n} \cdot {\bf s}_{j,n})
\right\rbrack 
\end{equation}
so that the total Hamiltonian of a collection of $M$ 
weakly interacting chains is
${\cal H}={\cal H}_{\rm inter}+
\sum_{i=1}^{M} ({\cal H}_{i,{\rm intra}})$.
Magnetic susceptibility and zero field specific heat 
measurements\cite{Affronte} were systematically performed 
on Gd(hfac)$_3$NITiPr: while no anomaly was found in the
magnetic susceptibility, a $\lambda$-type peak, like  
a second order phase transition, was clearly observed 
in the specific heat for $T\approx 2$K.
Moreover, a linear $T$-dependence was found for $C(T)$
at very low temperatures, inconsistently with
helical 3D long range order (which would give $T^3$).
This overall behavior was interpreted\cite{Affronte} 
in terms of a chiral phase transition for $T_{0}\approx 2$K
although the phase transition at $T_N$ to helical long range order, 
predicted by Villain,\cite{Statphys} was not observed down to the
lowest investigated temperature ($T_{\rm min}=0.175$K 
for Gd(hfac)$_3$NITiPr).
This was tentatively explained\cite{Affronte} taking into
account aging effects due to the moderate lability of the
organic radicals: even in very good and fresh samples, 
chains are reduced to segments of finite length and this has 
the effect of destroying the phase coherence necessary for 
the onset of 3D helical long range order.
On the contrary, chiral order is compatible
with the presence of finite segments since, for its onset,
it is only required that the chirality of the segments
is the same, a much less strict condition to be fulfilled.
In fact, while the $\lambda$-type anomaly in the specific heat
is very intense and robust and its position does not change
even in degradated samples, aging effects strongly influence
the magnetic susceptibility.\cite{Affronte}

In this paper, we report new specific heat measurements in 
magnetic field that demonstrate the magnetic nature of the 
phase transition at $T\approx 2$K. 
In order to obtain further information about the 
behavior of the two-spin correlation function, muon spin 
resonance ($\mu^+$SR) experiments in zero-field were also performed. 
Such a technique is particularly useful since it does not require 
the application of an external magnetic field, which in a sample 
with helical long range order is expected to cause the onset 
of complicated spin arrangements (``fan" phase).\cite{Coqblin}
To our knowledge, no $\mu^+$SR work on insulating 
helimagnets has been reported in the literature to date.

The paper is organized as follows. In Section II the new specific
heat data are reported, while in Section III the results of 
$\mu^+$SR experiments are shown. Finally, the conclusions are 
drawn in Section IV.

\section{Specific heat measurements}
Details about the preparation of samples can be found in 
Ref.~\onlinecite{Benelli}. The heat capacity of different pellets 
of pressed Gd(hfac)$_3$NITiPr micro-crystals was measured by using 
three different calorimeters and techniques: the adiabatic, the 
$ac-$ and the relaxation method. The heat signal was minimized in 
order to reduce the temperature ($T$) variation with respect to 
the  equilibrium state. The best results, on which we concentrate 
in the following, were obtained by using the relaxation method for 
which the $T$-modulation was less than 0.5\% of the base 
temperature. This implies an experimental resolution of 
$\log_{10}t \approx -2.5$ in terms of the reduced temperature 
$t=\vert T/T_0-1\vert$. Measurements in magnetic field were performed 
by using a Quantum Design PPM-7T System with a $^{3}$He insert and 
we begin presenting the new results of these experiments.

\begin{figure}
\label{fig1}
\centerline{
\includegraphics[width=8cm,angle=0,bbllx=38pt,%
bblly=251pt,bburx=556pt,bbury=615pt]{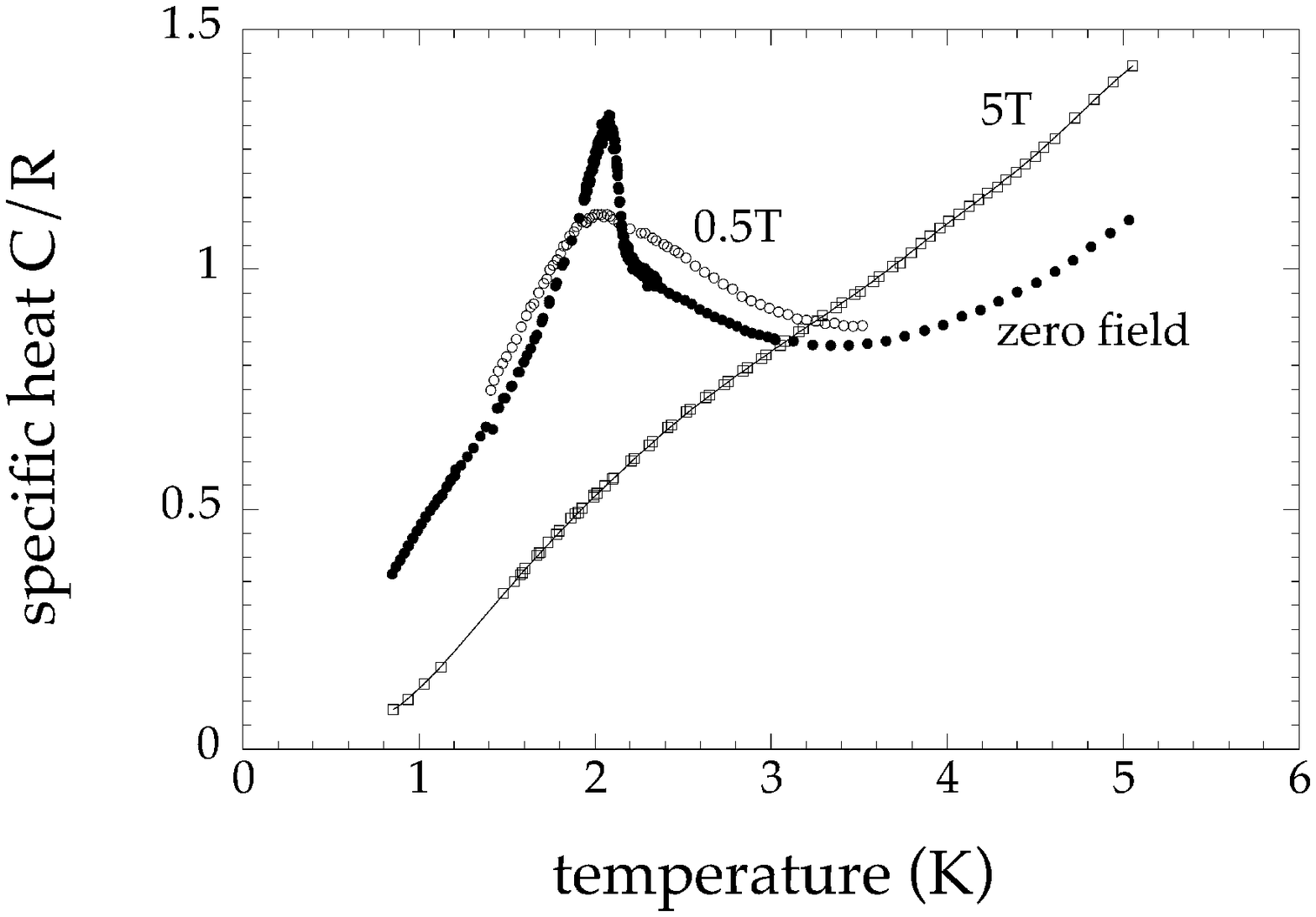}
}
\caption{The specific heat (normalized to the gas constant $R$)
versus temperature of Gd(hfac)$_3$NITiPr, 
measured in zero field (filled circles), in a field of 
0.5T (T=Tesla)(open circles), and 5T (open squares).}
\end{figure}

Fig.~1 shows the specific heat $C$, normalized to the gas constant
$R=8.314$J mol$^{-1}$ K$^{-1}$, as a function of temperature. 
In zero field a $\lambda$-anomaly is clearly visible at 
2.08$\pm$0.01K while, at lower temperatures, 
$C/R$ scales to zero with a quasi 
linear $T$-law. Note that the value of the transition temperature 
is reproducibly found in different samples (compare also data in 
Ref.~\onlinecite{Affronte}) in spite of their different quality and 
of the height of the $\lambda$-anomaly . The latter gets smoother 
in a magnetic field of 0.5T (T=Tesla) and is completely removed by the 
application of a 5T magnetic field. This definitively proves the 
magnetic origin of the transition, a fact that could not be surely 
deduced from results of previous experiments.\cite{Affronte}
Concerning the order of the phase transition, an important feature
is the absence of latent heat: we evaluated an upper limit of 
7x10$^{-3}$ J mol$^{-1}$. This fact seems to indicate that the phase 
transition is of the second order in zero field, although the
roundness of the peak, which can be due to the polycrystalline 
nature of the sample, prevents from determining the value of the 
critical exponent $\alpha$.\cite{notalog}
It is worth noticing that the rounded peak at 2K and the 
absence of a latent heat might be justified even in the case 
of a first-order chiral transition by the effect of microscopic
random quenched impurities\cite{Imry,Aize} that are surely present
in the magnetic chain.

In order to gain information about the magnetic order,
we determined the magnetic entropy removed in the low
temperature region $T<T_0$, and in particular that
below the anomaly. 
(The magnetic entropy for $T>T_0$ was not 
evaluated since for Gd(hfac)$_3$NITiPr the lattice contribution 
cannot be estimated in a reliable way owing to the complexity of 
the crystal structure and to the absence of a non-magnetic 
isomorph.) 
Such a calculation was performed using the specific heat data 
relative to sample C1 in Fig. 2 of Ref. \onlinecite{Affronte}, 
since it is considered the best sample hereto obtained, 
owing to its sensibly higher values of $C/R$. 
We found that the residual magnetic entropy below $T_0$ is 
rather large: 59\% of the total magnetic entropy, which 
is given by $R[\ln(2s+1)+\ln(2S+1)]=4R\ln 2$.\cite{nota}
Such a large value of the residual magnetic 
entropy below $T_0$ is probably due to a high concentration 
of defects that prevents the onset of a strong short range
order in the paramegnetic phase, as it usually happens 
in quasi one-dimensional
systems. For the sake of comparison, we remind that in the 
quasi one-dimensional antiferromagnet TMMC, the residual 
magnetic entropy below the N\'eel transition temperature 
$T_N$ was estimated to be nearly 
1\% of the total magnetic entropy.\cite{Borsa}
To obtain the entropy removed below the anomaly at $T_0$, 
one has to subtract the spin wave contribution (responsible for
the linear $T$-dependence of $C$ observed well below $T_0$)
from the total specific heat. In this way one obtains 
the value $\Delta S=0.59R$ for the magnetic entropy related to the
anomaly, in fair agreement with the value expected for 
an Ising system $\Delta S_{Ising}=R \ln 2\approx 0.69R$.
Such a result may be interpreted as a further indication for 
the onset, at $T_0=2.08$K, of a chiral transition which 
removes the Z$_2$ degeneracy between clockwise and 
anticlockwise chirality.

\section{$\mu^+$SR measurements}
The $\mu^+$SR data were collected in pellets of 
pressed Gd(hfac)$_3$NITiPr micro-crystals 
at the ISIS facility (Rutherford Appleton Laboratory, UK), 
in the temperature range 0.3-300 K, in zero magnetic field
(for data in longitudinal applied field, see  
Ref.~\onlinecite{lascialfari}).
After a proper background (titanium sample holder) 
subtraction, the muon asymmetry A (whose total value is A$\sim$0.21, 
close to 0.23, the expected value on the EMU beamline at RAL), 
can be analyzed by means of 
a single "stretched" exponential decay $e^{-(t/ \lambda)^{0.5}}$ 
in the whole $T$-range (see Fig.~2).

\begin{figure}
\label{fig2}
\centerline{
\includegraphics[width=8cm,angle=90,bbllx=0pt,bblly=0pt,%
bburx=556pt,bbury=615pt]{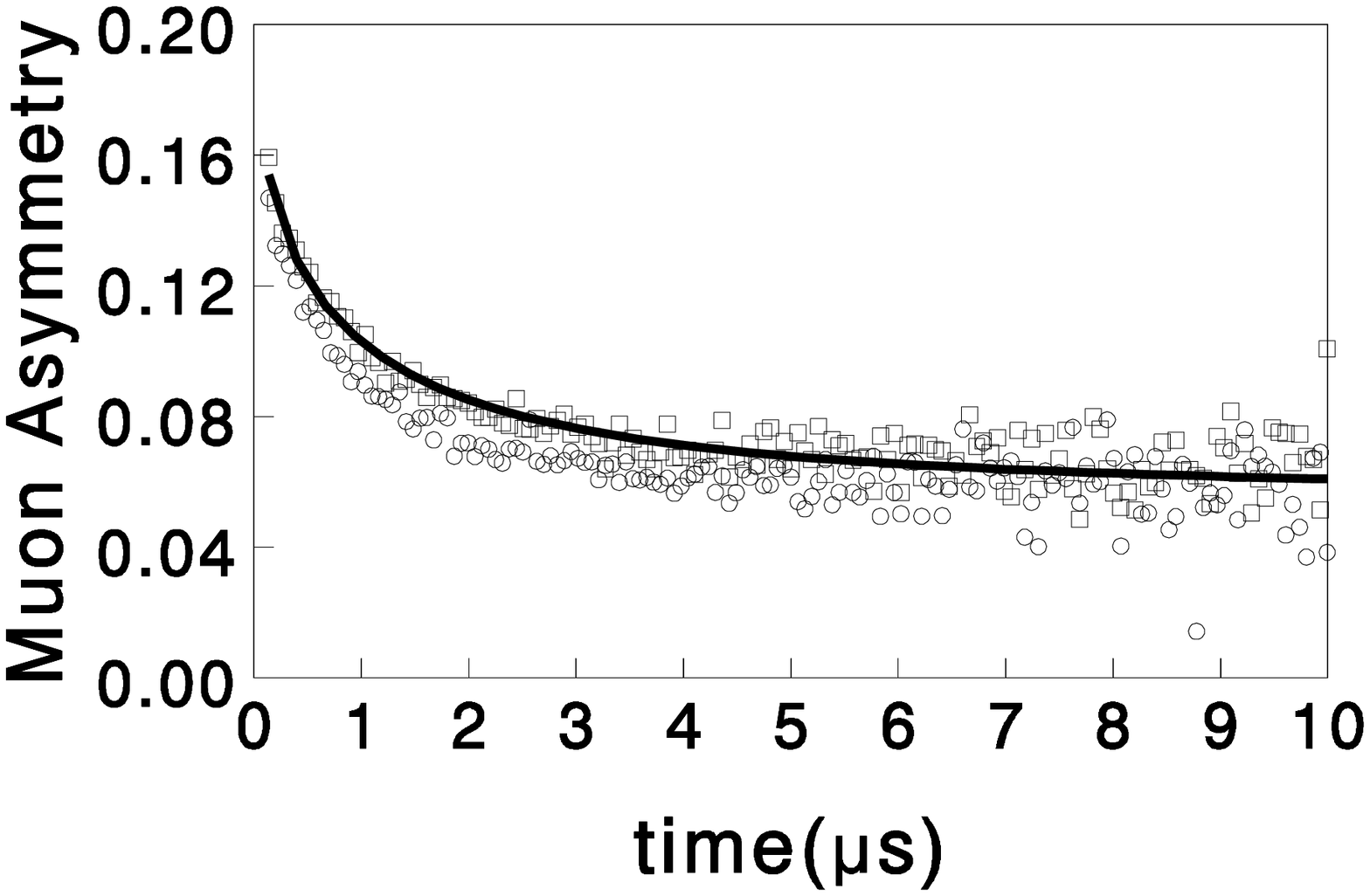}
}
\caption{Muon asymmetry at two different temperatures. The solid
line is a fit of the $T$=4K data (open squares), obtained by 
means of a stretched exponential.
As can be evinced from the figure, the stretched 
behavior is also followed 
by the data at $T$=0.3K (open circles). 
}
\end{figure}

This fact implies, in the whole investigated $T$-range : 

(i) a distribution of relaxation rates $\lambda$, witnessed by
the stretched behavior;

(ii) the absence of a local field 
due to long-range magnetic order at the muon site 
(in presence of long-range magnetic order, 
one should observe\cite{schenck}
an  oscillation in the asymmetry A, caused 
by the precession of the muon spin around the 
local field itself, or a loss of asymmetry) ; 

(iii) as in the case of short-range magnetic order the muon asimmetry
should display a Kubo-Toyabe-like behavior,\cite{schenck} one guesses a
wide distribution of local fields values at the muon site,
further broadened by powders average.

Further support for the absence of long-range order 
comes from the analysis of the
temperature behavior of the muon longitudinal 
relaxation rate $\lambda$ (see Fig.~3).

\begin{figure}
\label{fig3}
\centerline{
\includegraphics[width=8cm,angle=0,bbllx=1pt,bblly=64pt,%
bburx=590pt,bbury=755pt]{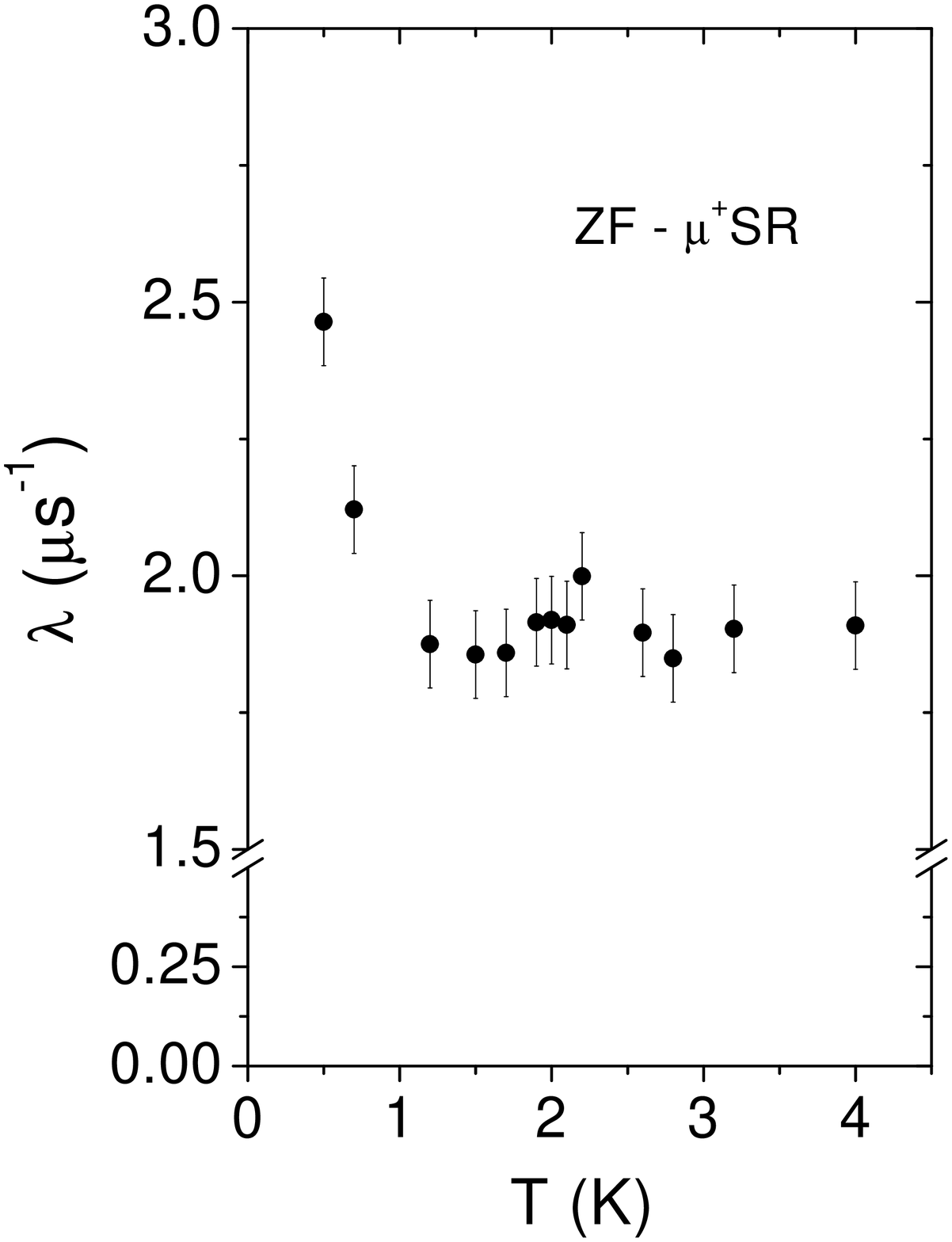}
}
\caption{Muon longitudinal relaxation rate 
measured in Gd(hfac)$_3$NITiPr
as a function of temperature
in zero applied magnetic field. 
}
\end{figure}

The parameter $\lambda$ represents the muon 
relaxation rate as a result of the muon-lattice interaction. 
The general formula for $\lambda$ in a weak collision approach 
for fluctuations faster than the muon resonance frequency,
$\omega_R$, is\cite{borsariga}
\begin{equation}
\label{lambda}
\lambda \propto \gamma^2[J_+ (\omega_R) + J_-(\omega_R)] 
\end{equation}
where
\begin{equation}
J_\pm = \int  d t~e^{-i\omega_R t} 
\langle
{\bf h_\pm}(t)\cdot {\bf h_\pm}(0)
\rangle
\end{equation}
are the spectral densities, at the resonance frequency $\omega_R$, 
of the correlation functions for the local hyperfine 
(dipolar and contact) field components $h_\pm$. Let us now
assume that the electronic spin components $S_\alpha$ 
(coming from the Gd ion) and $s_\alpha$ (coming from the iPr 
organic radical) have isotropic fluctuations.

To analyze the muon relaxation rate behavior in terms of scaling 
laws, we consider the three types of correlation functions 
that involve two spins.\cite{Affronte}
Thus, Eq.~(\ref{lambda}) can be rewritten in terms of the
collective spin components ${\bf S_q}$ and ${\bf s_q}$ as
\begin{eqnarray}
\label{collective}
\lambda &\propto& \gamma^2 {1\over N}
\sum_{i,j=1,2} \sum_{\bf q} \vert{{\bf h}_{\bf q}}\vert ^2
\int dt ~
\langle
\roarrow{\sigma}_{i{\bf q}}^{+} (0) 
\roarrow{\sigma}_{j{\bf q}}^{-} (t)
\rangle 
\cr 
&=&
\gamma^2 {1\over N} \sum_{i,j=1,2} 
\sum_{\bf q} 
\vert{{\bf h}_{\bf q}}\vert ^2
\vert \roarrow{\sigma}_{i{\bf q}}\vert
\vert \roarrow{\sigma}_{j{\bf q}}\vert
{1\over {\Gamma_{\bf q}}}
\end{eqnarray}
where $\roarrow{\sigma}_1={\bf S}$ and $\roarrow{\sigma}_2={\bf s}$;  
$\Gamma_ {\bf q}$ is the decay rate of the collective spin 
fluctuations, while ${\bf h}_{\bf q}$ is the Fourier transform 
of the lattice functions that couple the muons to the spins of 
the magnetic ions. Now, we will show that, in the event that 
long range order sets in below a transition temperature 
$T_{\rm tr}$, then a divergence in the muon longitudinal 
relaxation rate $\lambda$ is expected (at least for $H=0$) 
upon approaching $T_{\rm tr}$ from higher temperatures.
In fact, by expanding the ${{\bf h}_{\bf q}}$ vector around 
the value corresponding to the critical wave vector 
${\bf Q}$ (which characterizes the long range order below 
the transition temperature $T_{\rm tr}$), and by using conventional
scaling arguments for the $\bf q$-dependence of the correlated
fluctuations,\cite{borsariga} Eq.~(\ref{collective}) takes the form
\begin{equation}
\lambda \propto \sum_i^3 (\gamma h^i_{\rm eff})^2 
{1\over N} \sum_{\bf q_i}
\frac{\xi_i^{2-\eta_i}f(q_i\xi_i)}
{\omega_e^i \xi_i^{-z_i}g(q_i\xi_i)}
\end{equation}
where $h^i_{\rm eff}$ are the static fields at the muon sites, 
resulting from the ordered configuration of the Gd and 
radical spins, while $\bf q_i$ are the wavevectors
measured with respect to the critical ones; $\xi_i$ are the 
correlation lengths; $\omega_e^i$ are the Heisenberg exchange 
frequencies.
The decay rate of two-spin fluctuations is given by
$\Gamma_{q_i}=(\omega_e / \xi_i^z)g(q_i\xi_i)$, 
where $g(q_i\xi_i)$ and $f(q_i\xi_i)$ are homogeneous
functions of the product $x=q_i\xi_i$. 
By transforming the {\bf q}-summation into an integral, 
and noting that the latter converge to a number of the 
order of unity, one finally obtains
\begin{equation}
\lambda \propto \sum_{i=1}^3~ (\gamma h^i_{\rm eff})^2 ~
{1\over  { \omega_e^i}}~
\xi_i^{z_i-d+2-\eta_i}
\end{equation}
By setting $\xi_i\propto (T-T_{\rm tr})^{-\nu_i }$, 
a divergence of the form
\begin{equation}
\lambda \propto \sum_{i=1}^3~ (T-T_{\rm tr})^{-n_i}
\end{equation}
with $n_i=\nu_i(z_i-1-\eta_i)$, is expected.
As can be easily seen from Fig.~3, the muon longitudinal 
relaxation rate $\lambda$ does not present any 
divergence at $T\approx 2$ K for Gd(hfac)$_3$NITiPr, 
thus indicating that this temperature does not mark 
any transition $T_{\rm tr}$ to a long range order 
for the two-spin correlation function.

Finally we observe that one of the main motivations of the present
$\mu^+$SR measurements was that of investigating 
the behavior of the muon longitudinal relaxation rate 
$\lambda$ around the temperature for which 
the specific heat presents a peak and that 
the main result of the $\mu^+$SR investigation presented
in this paper is the absence of any anomaly in $\lambda$ 
at $T_0 \approx 2$ K.
The strong increase of $\lambda$ observed below 1 K 
may have different possible explanations: {e.g.}, a progressive 
increase of the coherence lengths $\xi_i$'s or a resonance 
with a spin-wave mode. Further measurements of $\lambda$ 
for temperatures lower than 1K should be necessary in 
order to ascertain the origin of such a feature, but this 
goes beyond the scope of the present work.

\section{conclusions}
In conclusion, we have performed specific heat and 
$\mu^+$SR experiments in the molecular magnetic chain 
Gd(hfac)$_3$NITiPr. While the four-spin correlation function 
(probed by the magnetic specific heat) showed 
a seemingly continuous magnetic phase transition 
at $T \approx 2$K, {\it the two-spin correlation function}
(probed by $\mu^+$SR) {\it did not show any anomaly related to 
a transition to conventional long range helical order} 
(similar to that present in the 3D helimagnets Tb, Yb, 
Dy, for instance).\cite{Kawamura}
Moreover,  below  $T \approx 2$K 
no precession of the muon spin or loss of 
muon asymmetry similar to what happens in the case of 
long-range magnetic order, was observed.
Taking also into account that at very low temperatures 
(0.175-0.8 K) a linear $T$-dependence of the magnetic 
specific heat was observed in Gd(hfac)$_3$NITiPr (a clear 
indication for isotropic 1D magnetic behavior),\cite{Affronte} 
and relying upon an estimation of the residual 
magnetic entropy below the anomaly,
we are led to conclude that, below $T_0=2.08$ K, 
an ordered chiral phase without 3D long range helical 
order sets in.
\newline

\begin{acknowledgments}
This work was partially funded by PRA99-INFM, EU Network ``3MD" 
(ERB 4061, PL 97-0197), PRIN2001 of the Italian MIUR.
We gratefully acknowledge J. Villain for fruitful discussions 
and suggestions.
\end{acknowledgments}



\begin{references}
\bibitem[9]{notalog}
In fact, the quantity $\log_{10}(\Delta C/R)$-vs-$\log_{10}t$,
where $\Delta C$ is the fluctuation contribution to the 
specific heat, is experimentally found to saturate 
and become $t$-independent for $\log_{10}t \le -2$.
\bibitem[12]{nota} 
It is worth noting that nearly 50\% of the residual magnetic 
entropy below $T_0$ is due to to the contribution of the free spin waves 
of the one-dimensional system (which are responsible for
the linear $T$-dependence of the magnetic specific heat
at very low temperatures).
\end{references}
\end{document}